\documentclass[final,5p,twocolumn,lettersize]{elsarticle}

\usepackage{graphicx}
\biboptions{sort&compress}

\usepackage{amssymb}

\newcommand{\cobaltite}{Co$_3$O$_4$}
\newcommand{\spinel}{MgAl$_2$O$_4$}
\newcommand{\sapphire}{$\alpha$-Al$_2$O$_3$}

\journal{Journal of Crystal Growth}

\begin{document}

\begin{frontmatter}



\title{Growth and characterization of thin epitaxial {C}o$_3${O}$_4$(111) films}

\author[addr1,addr0]{C.~A.~F.~Vaz\corref{cor1}} \cortext[cor1]{Corresponding author. Email: carlos.vaz@yale.edu}%
\author[addr1,addr0]{V.~E.~Henrich}
\author[addr1,addr0]{C.~H.~Ahn}
\author[addr2,addr0]{E.~I.~Altman}

\address[addr1]{Department of Applied Physics, Yale University, New Haven, Connecticut 06520}%
\address[addr2]{Department of Chemical Engineering, Yale University, New Haven, Connecticut 06520}%
\address[addr0]{Center for Research on Interface Structures and Phenomena (CRISP), Yale University, New Haven, Connecticut 06520}%

\begin{abstract}
The growth and characterization of epitaxial \cobaltite(111) films
grown by oxygen plasma-assisted molecular beam epitaxy on single
crystalline \sapphire(0001) is reported. The \cobaltite(111) grows
single crystalline with the epitaxial relation
\cobaltite(111)[$\bar{1}$2$\bar{1}$]$\parallel$\sapphire(0001)[10$\bar{1}$0],
as determined from {\it in situ} electron diffraction. Film
stoichiometry is confirmed by x-ray photoelectron spectroscopy,
while {\em ex situ} x-ray diffraction measurements show that the
\cobaltite\ films are fully relaxed. Post-growth annealing induces
significant modifications in the film morphology, including a
sharper \cobaltite/\sapphire\ interface and improved surface
crystallinity, as shown by x-ray reflectometry, atomic force
microscopy and electron diffraction measurements. Despite being
polar, the surface of both as-grown and annealed
{C}o$_3${O}$_4$(111) films are ($1\times 1$), which can be explained
in terms of inversion in the surface spinel structure.
\end{abstract}


\end{frontmatter}


\section{Introduction}

Recent developments in the growth of high quality epitaxial thin
metal oxide films have led to a renewed interest in the properties
of metal oxides as one or more physical dimensions is reduced to the
nanoscale. As the contribution of the interface becomes a
significant part of the whole system, new physical phenomena are
expected to emerge due to symmetry-breaking and the ensuing changes
in electronic structure. Surface states and perturbed orbital states
of the interface atoms often give rise to strongly anisotropic
behavior and novel surface effects. Critical to this effort is the
ability to grow and characterize high quality epitaxial thin films.
It is in this context that we report here a detailed study of the
growth and structural and electronic characterization of epitaxial
[111]-oriented \cobaltite\ thin films.

Of the three known cobalt oxides, the mixed valence compound,
Co$^{2+}$Co$_2^{3+}$O$_4$, is stable at ambient pressure and
temperature and crystallizes in the cubic spinel structure (with
lattice constant $a=8.086$ \AA\ \cite{PBBC80}), while  the high
temperature CoO phase crystallizes in the rock salt structure. Also
reported in the literature is a sesquioxide of cobalt, Co$_2$O$_3$,
crystalizing in the corundum structure \cite{CJM71,Samsonov82} ({\it
ab initio} calculations suggest this phase to be a stable energy
minimum \cite{CS97}). One aspect of particular interest in compounds
with strong ionic character is the effect of surface charge on the
stability of polar surfaces and interfaces. This occurs along
crystal directions where an electric dipole moment is present
(arising from alternating charged crystal planes), where a divergent
electrostatic energy would develop in clean, bulk-terminated
crystals. One general mechanism for quenching such an increase in
electrostatic self energy is via charge compensation, whereby a
modification in the surface charge distribution cancels the overall
electric dipole \cite{Tasker79,Noguera00,GFN08}. Charge compensation
is bound to result in important modifications of the surface atomic
and electronic structure, including changes in the valence state of
surface ions, surface reconstructions, surface roughening and
faceting, among others
\cite{HKWG78,Tasker79,Noguera00,JPS+02,LPP+05,GFN08}. In the case of
the spinel structure, all low index surfaces are polar, and we
expect charge compensation processes to modify the atomic and
electronic surface structure of spinel crystals. Recently, we have
shown that \cobaltite(110) thin films grown by molecular beam
epitaxy (MBE) exhibit ($1\times 1$) surfaces, and we attributed the
stability of this surface structure to modified cationic valence
states at the surface, a process equivalent to an inversion in the
spinel structure at the film surface  \cite{VWA+09}. Motivated by
these findings, we study here the (111) surface of \cobaltite\ grown
on \sapphire(0001) single crystals.

\begin{figure}[t!bh]
\begin{centering}
\includegraphics*[width=\columnwidth]{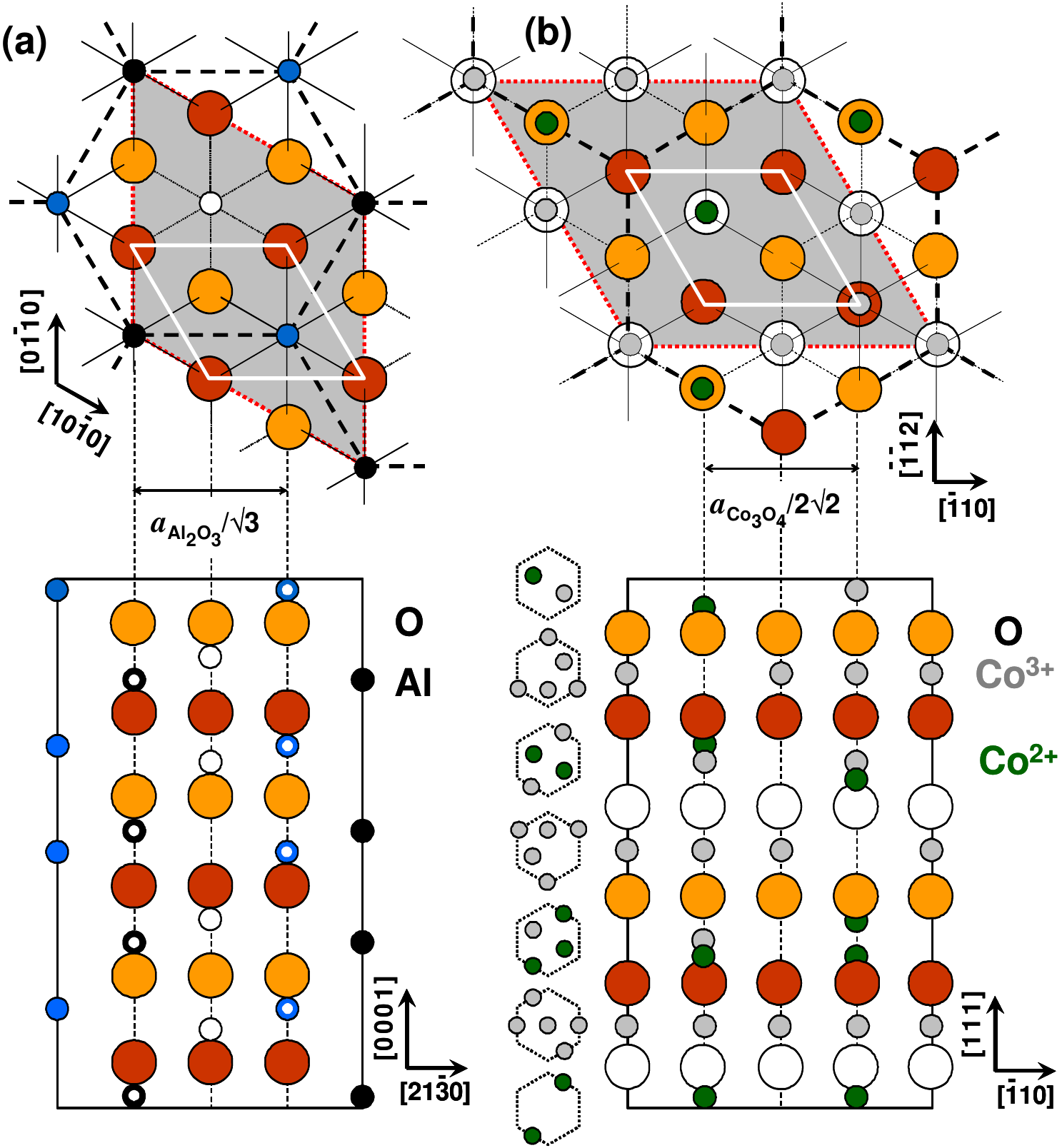}
\caption{(a) Schematic diagram of the crystal structure of corundum
(\sapphire), showing the primitive hexagonal (oblique) cell of the
(0001) surface (shadowed area, top) and the stacking sequence along
the [0001] direction in the hexagonal motive (bottom). Large circles
stand for O atoms and small circles for Al: white small circles
correspond to Al atoms lying in the center of the hexagonal motive,
grey and dark circles to Al atoms lying at the vertices. White dots
on the Al atoms designate positions where two atoms overlap along
the direction perpendicular to the plane of the page. (b) Schematic
diagram of the crystal structure of the spinel \cobaltite\ for the
primitive oblique cell of the (111) surface in the hexagonal
representation (shadowed area, top) and the respective atomic
stacking along the [111] direction (bottom). Large circles represent
O atoms, small circles correspond to Co (light circles are
Co$^{2+}$, tetrahedrally coordinated; darker circles are Co$^{3+}$,
octahedrally coordinated). To the left, in the bottom diagram, is
shown the cationic arrangement of the (111) planes along the [111]
direction. The oblique surface unit cell of the O sublattices are
drawn in white (top).} \label{fig:surface}
\end{centering}
\end{figure}

The crystal structure of \cobaltite\ along the [111] direction is
particularly intricate: while the hexagonal primitive (oblique) cell
in the (111) plane is relatively small, with a lattice constant of
5.72 \AA\ (see Fig.~\ref{fig:surface}b), the repeat unit along the
[111] direction contains 18 atomic planes, in the form
[O$_4$-Co$^{3+}_3$-O$_4$-Co$^{2+}$-Co$^{3+}$-Co$^{2+}$]$_3$, with
four basic types of planes: one hexagonal oxygen plane, two
octahedral Co$^{3+}$ planes and one tetrahedral Co$^{2+}$ plane.
Along the [111] direction, the O sublattice in \cobaltite\ presents
a face-centered cubic close-packing (fcc) stacking sequence, or
A-B-C-A. All planes have non-zero charge per unit cell, and
therefore all (111) planes are polar; since the (111) planes are
composed of only O anions or Co cations, the total charge per unit
area is large.

Very few studies have addressed the stability of the \cobaltite(111)
surface. Meyer et al.~\cite{MBG+08} have reported the growth of
twined ($1\times 1$) [111]-oriented \cobaltite\ films on
Ir(001)-($1\times 1$); from scanning tunneling microscopy and
quantitative low energy electron diffraction (LEED) analysis, they
conclude that the film surface is terminated at a Co$^{2+}$-O plane,
with a strong inward relaxation of the Co atoms to almost level with
the O plane, and they suggest that a modified ionicity (inversion)
of the surface cations leads to charge compensation and
stabilization of the ($1\times 1$) \cobaltite(111) surface. Other
recent studies of the (111) surface of \cobaltite\ include that of
Petitto et al.~\cite{PMCL08}, where a detailed study of the
interaction of the low index surfaces of \cobaltite\ to oxygen and
water is reported, and the study by Tang et al.~\cite{TLH08} on the
reactivity of \cobaltite\ nanoparticles, which is found to be
strongly reduced for particles terminated by \{111\} facets, as
compared to irregularly shaped nanoparticles. Studies of the surface
energies of \cobaltite\ could not be found in the literature. As a
proxy system, we may consider the case of \spinel, another
prototypical spinel with a nearly identical lattice constant, which
has been studied more often. Theoretically, atomistic calculations
of the surface energy of \spinel\ for the low index planes indicate
that charge compensation and surface stability can be achieved by
surface vacancies; for the (110) and (111) surfaces, significant
surface relaxations are predicted. For the (111) surface, the lowest
energy termination is that which truncates the crystal at the O
layer between the Al$^{3+}$ and the Al$^{3+}$-Mg$^{2+}$-Al$^{3+}$
layers, with nine O per unit cell on top of the Al layer, and seven
O on the Mg layer at the opposite face of the crystal \cite{FPW00}.
An earlier study predicted that the lowest energy (111) surface is
that between the Mg$^{2+}$ and Al$^{3+}$ planes, with two vacant
Al$^{3+}$ cations per unit cell; the possibility of surface
inversion was also considered and calculated to lower the surface
energy \cite{DPW94}. While such inversion may be chemically
difficult to achieve in \spinel\ during cleavage, it could be
produced during crystal growth \cite{GFN08}. In the case of
\cobaltite, as noted, the mixed Co valency may make this process
more easy to accommodate, even in bulk crystals, since it does not
involve atomic diffusion. Cleavage of \spinel\ has been shown to
occur preferentially along (001) planes, which also exhibit the
lowest fracture surface energy \cite{SB80,Bradt97}. These results
suggest that the (100) surface has the lowest energy, which is in
disagreement with the results of the most recent theoretical
calculations, which indicate the compensated (111) surface to be the
most stable; this discrepancy between the fracture experimental
results and the atomistic calculations has been attributed to the
effect of water adsorption, which was found to reduce significantly
the free energy of the (111) surface \cite{FWP01,LDL+05,GFN08}.

Sapphire, \sapphire, crystalizes in the corundum structure and is
the most stable aluminum oxide. \sapphire\ is rhombohedral (with two
formula units in the primitive cell), but it is more conveniently
described in the hexagonal representation, with lattice parameters
$a = 4.7570$ \AA, $c = 12.9877$ \AA\ \cite{NH62,Cousins81,KE90} for
the primitive (oblique) hexagonal cell, shown in
Fig.~\ref{fig:surface}a following Wyckoff's convention for the
crystal axes \cite{Wyckoff64}. The hexagonal primitive cell contains
six formula units, with a stacking sequence of eighteen layers, in
the form [Al-O$_3$-Al]$_6$, along the $c$-axis. This results in
three possible bulk terminations of the unit cell along the [0001]
direction, consisting of planes exposing Al atoms for a truncation
between two Al atomic layers, planes exposing two Al planes (Al-Al),
and planes exposing O atoms, see Fig.~\ref{fig:surface}a
\cite{GEL92,GL94}; the Al termination yields non-polar surfaces.
Experimentally, it is observed that (0001) surfaces prepared at
temperatures below $\sim$1500 K show a stable ($1 \times 1$)
surface, corresponding to the non-polar Al-termination; this surface
is also characterized by large inward relaxation of the Al atoms
\cite{MVG93,AR97,GRBG97,Renaud98,JC01,WAM+06}. Annealing to higher
temperatures produces a rotated ($\sqrt{31} \times \sqrt{31}$)
surface reconstruction, associated with the Al-rich (Al-Al) surface
\cite{Chang68,FS70,HC94,Renaud98,SHOS99,WMSH00,BR01,SHWM02,MP04,Kelber07}.
Figure~\ref{fig:surface}a shows the bulk terminated \sapphire(0001)
surface exposing an Al-Al plane. The O sublattice is close to a
hexagonal close packed (hcp) lattice (A-B-A stacking, with the Al
atoms occupying 2/3 of the octahedrally coordinated interstitial
sites), with an O inter-plane distance of 2.166 \AA\ along the
[0001] direction.

Here, we consider the surface and interface properties of
[111]-oriented epitaxial films of the prototypical \cobaltite\
spinel grown on \sapphire(0001) substrates. Since the oxygen
sublattices of both \sapphire\ and \cobaltite\ are close packed
(hexagonal and face-centered cubic, respectively) with close basal
lattice constants, 2.746 \AA\ for \sapphire\ and 2.858 \AA\ for
\cobaltite\ (lattice mismatch of --3.9\%), we may expect epitaxial
growth of \cobaltite\ on \sapphire(0001) to proceed as a
continuation of the O-sublattice. Indeed, we show that
\cobaltite(111) thin films can be grown epitaxially on
\sapphire(0001) substrates by oxygen assisted molecular beam
epitaxy. The as-grown \cobaltite(111) surface exhibits a ($1\times
1$) surface diffraction pattern with no evidence of periodic
reconstructions. The as-grown film surface shows a significant
amount of disorder, which is slightly reduced upon annealing in air.

\section{Sample growth and characterization techniques}

The samples for this study were grown by O-assisted molecular beam
epitaxy (MBE) in a ultrahigh vacuum MBE growth system (base pressure
of $1 \times 10^{-9}$ mbar). The substrates consist of polished
\sapphire(0001) single crystal wafers, which were annealed at 870 K
in ultrahigh vacuum for 60 min, and then cleaned under an O-plasma
flux at 470 K for 30 min prior to film growth. No impurities other
than trace amounts of Ca were detected by Auger electron spectrocopy
(AES) taken after the substrate cleaning procedure. Low energy
electron diffraction (LEED) and reflection high energy electron
diffraction (RHEED) of the \sapphire(0001) substrate after cleaning
display patterns characteristic of highly ordered surfaces (see
Figs.~\ref{fig:RHEED} and \ref{fig:LEED}), indicating a good
crystalline order of the surface. The LEED and Laue diffraction
patterns show 3-fold symmetric patterns (inner ring of the six
diffraction spots in the LEED pattern show alternating intensity),
suggesting that the surface is composed, predominantly, of
double-layer atomic steps in the single crystalline \sapphire(0001)
surface. The cobalt oxide films were grown by oxygen-assisted
molecular beam epitaxy by simultaneous exposure of the substrate to
a thermally evaporated Co atomic beam and an atomic O flux. The
oxygen partial pressure during growth was $3\times 10^{-5}$ mbar,
and the electron cyclotron resonance oxygen plasma source magnetron
power was set to 175 W, yielding an atomic O flux of the order of
$1\times 10^{14}$ cm$^{-2}$s$^{-1}$ at the sample
\cite{GKA05,AWN+00}. The Co evaporation rate was $\sim$2 \AA/min,
estimated by means of a calibrated quartz crystal microbalance.
Sample growth was monitored with RHEED, and film crystallinity and
electronic structure were determined immediately after growth by
RHEED, LEED and XPS, entirely in ultrahigh vacuum. After film
growth, the samples were characterized {\it ex situ} using x-ray
diffraction (XRD), x-ray reflectometry (XRR), and atomic force
microscopy (AFM). Post-growth annealing was performed at 820 K for
14 h in air; this temperature and oxygen partial pressure favor the
formation of \cobaltite\ over CoO \cite{TK55,KY81a,OS92a}. For this
study, two \cobaltite\ samples (22 and 38 nm thick) were grown
independently, which were found to have similar spectroscopic,
structural and morphological characteristics (due to sample
charging, no LEED patterns could be obtained for the annealed 22 nm
film and for the as-grown 38 nm film).

\section{Results and discussion}

\begin{figure}[t!bh]
\begin{centering}
\includegraphics*[width=\columnwidth]{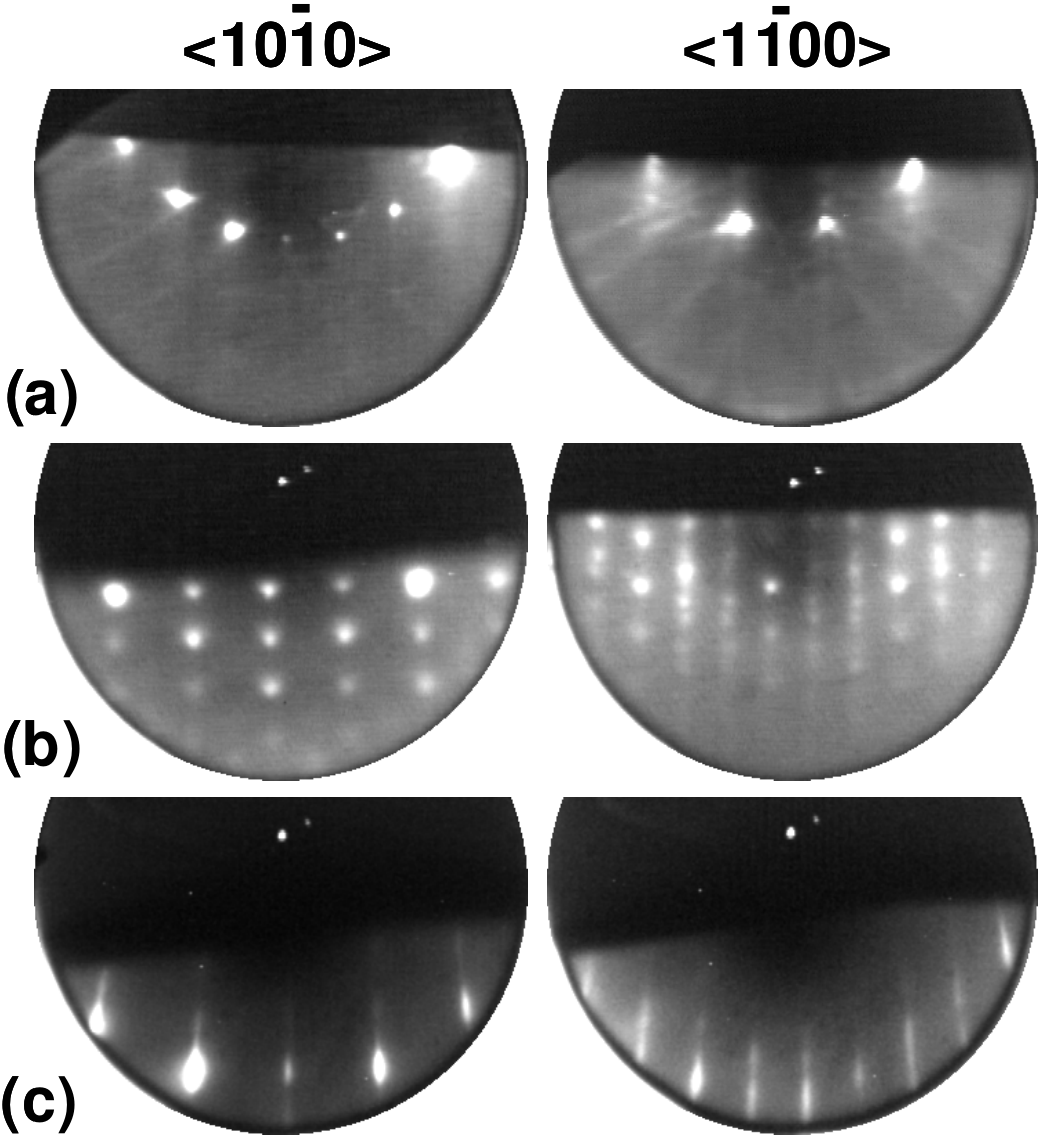}
\caption{Reflection high energy electron diffraction (RHEED)
patterns of (a) the \sapphire(0001) substrate and \cobaltite\ films
(b) before and (c) after annealing, along the $\langle
10\bar{1}0\rangle$ and $\langle 1\bar{1}00\rangle$ azimuths of the
\sapphire\ crystal (parallel to the electron beam, set at a grazing
angle of incidence). Patterns (a) and (b) are from the 22 nm sample,
(c) is from the 38 nm \cobaltite\ sample. The incident beam energy
was set to 15 keV.} \label{fig:RHEED}
\end{centering}
\end{figure}

\begin{figure}[t!bh]
\begin{centering}
\includegraphics*[width=\columnwidth]{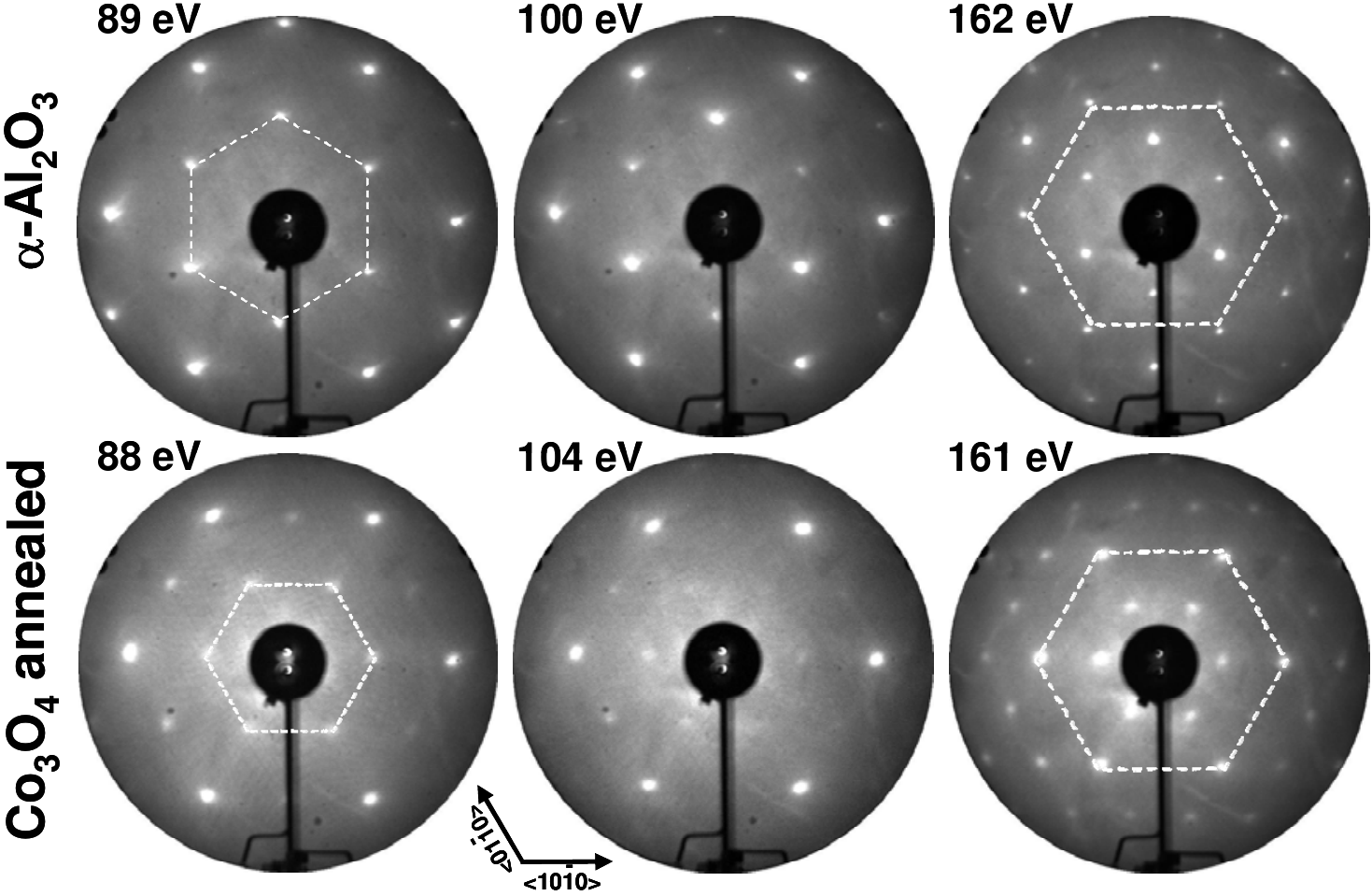}
\caption{Low energy electron diffraction (LEED) patterns of the
\sapphire(0001) (top panel) and for the annealed 38 nm \cobaltite\
film (bottom panel) at three values of the incident beam energy. The
in-plane \sapphire(0001)\ crystal orientation, inferred from Laue
diffraction, is shown in the inset. The dashed hexagon indicates the
symmetry of the LEED pattern; based on the geometrical configuration
of our LEED system, the hexagon on the left images correspond to the
respective unit cells of \sapphire\ and \cobaltite, while the larger
hexagons on the right correspond to the hexagonal unit cell of the O
sublattice, identical for both \sapphire\ and \cobaltite.}
\label{fig:LEED}
\end{centering}
\end{figure}

The surface crystallinity of the films was probed during growth by
RHEED. The RHEED pattern evolution showed a gradual fading of the
\sapphire\ sharp diffraction spots with increasing Co oxide
thickness, became streaky at about 15 \AA\ and finally broadened at
about 30 \AA. This indicates that film growth occurs via
three-dimensional island growth. Typical RHEED patterns of the
\cobaltite\ films after growth are shown in Fig.~\ref{fig:RHEED},
where the in-plane crystal directions refer to those of the
\sapphire\ substrate, determined independently from Laue diffraction
measurements. The RHEED patterns exhibited relatively broad
diffraction spots, suggesting a relatively rough surface. The
diffraction patterns are characteristic of a transmission pattern of
the spinel \{112\} planes along the $\langle 10\bar{1}0\rangle$
azimuths of the \sapphire\ substrate (similar to the diffraction
pattern of as-grown \cobaltite(110) films along the [112] direction
observed in \cite{VWA+09}, but rotated by 90$^\mathrm{o}$); and of
the spinel \{110\} planes along the $\langle 1\bar{1}00\rangle$
azimuths. In particular, we infer the epitaxial growth relation as
\cobaltite(111)[$\bar{1}$2$\bar{1}$]$\parallel$\sapphire(0001)[10$\bar{1}$0]
and
\cobaltite(111)[01$\bar{1}$]$\parallel$\sapphire(0001)[1$\bar{1}$00].
The same epitaxial relationship is observed for
\spinel/\sapphire(0001) grown by solid state reactions
\cite{RF63,LCC05}. LEED patterns for the \sapphire(0001) substrate
(Fig.~\ref{fig:LEED}) and for the as-grown \cobaltite\ film (not
shown) exhibit a ($1\times 1$) diffraction pattern. Compared with
the \sapphire\ LEED patterns, the diffraction spots of the as-grown
\cobaltite\ film are much broader, and the background is also more
intense, indicating that the as-grown \cobaltite\ films have a
significant amount of surface disorder (charging of the surface also
contributes to the poor patterns, especially at lower electron beam
energies). Motivated by our recent results that demonstrated a
significant improvement in the bulk and surface structure of
\cobaltite/\spinel(110) thin films upon annealing \cite{VWA+09}, we
have also studied the effect of post-growth annealing on the
properties of \cobaltite/\sapphire(0001). RHEED patterns obtained
after annealing (shown in Fig.~\ref{fig:RHEED} for the 38 nm film)
show that annealing induces significant transformations in the film
structure, as indicated by sharper and streakier RHEED diffraction
patterns. However, the RHEED characteristics indicate that these
surfaces are not atomically flat. LEED patterns for the annealed
sample show well defined ($1\times 1$) patterns with sharp
diffraction spots; although the patterns are not as sharp as for the
\sapphire\ substrate and while the background is more intense, it
can be inferred that the \cobaltite\ film surface is well ordered.
Unlike the 3-fold symmetric \sapphire\ LEED patterns, the LEED
diffraction patterns of the \cobaltite\ are 6-fold symmetric; this
is usually associated with the presence of rotational twinning,
i.e., the presence of both ABCA and ACBA stakings, which is known to
occur in the growth of fcc (111) films on the (0001) planes of
hexagonal crystals \cite{Stowell75,GKCB97}. Based on the geometrical
configuration of our LEED system we are able to deduce the
reciprocal space unit cell of the \cobaltite(111) and
\sapphire(0001) surfaces (Fig.~\ref{fig:LEED}, left) and that of the
respective O sublattices (Fig.~\ref{fig:LEED}, right). It is readily
seen that the unit cell of the \sapphire(0001) surface is larger (in
reciprocal space) than that of \cobaltite(111) and rotated by
30$^\mathrm{o}$, while the surface unit cell of the O sublattice are
juxtaposed one on the other as expected from growth of \cobaltite\
as a continuation of the O sublattice. The relative spacing of the
diffraction spots in the \sapphire\ and \cobaltite\ RHEED patterns
along the direction perpendicular to the electron beam also agree
with this structural model: the spacing ratio is $\approx 0.7$ along
the $\langle 10\bar{1}0\rangle$ azimuth and $\approx 2.0$ along
$\langle 1\bar{1}00\rangle$, which correspond to
$\sqrt{3}a_\mathrm{Co_3O_4}/a_\mathrm{\alpha-Al_2O_3} = 0.69$ and
$a_\mathrm{Co_3O_4}/\sqrt{3}a_\mathrm{\alpha-Al_2O_3} = 2.08$,
respectively.

\begin{figure}[t!bh]
\begin{centering}
\includegraphics*[width=\columnwidth]{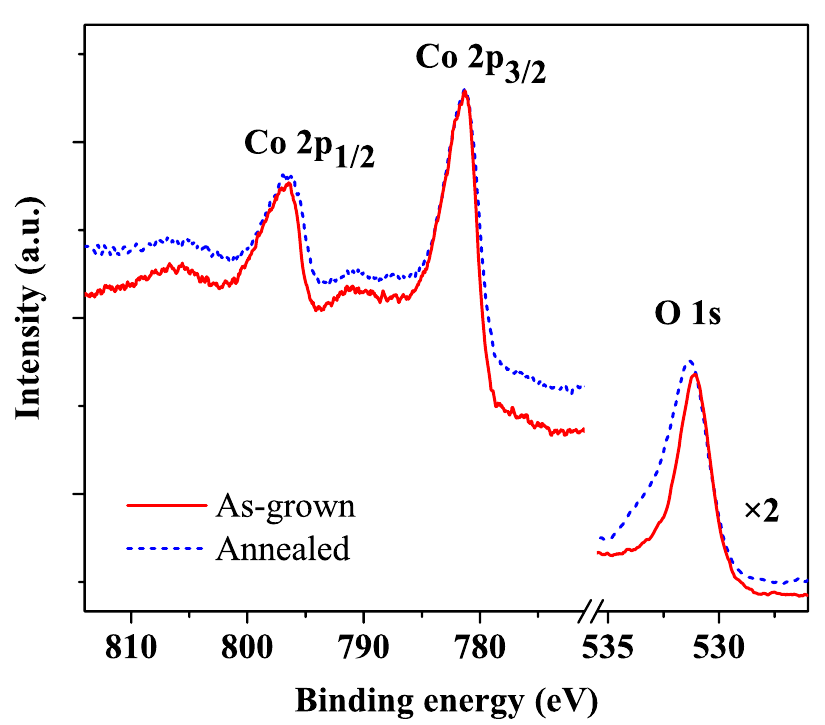}
\caption{X-ray photoelectron spectra for as-grown and annealed 22 nm
\cobaltite\ film for the Co 2p and O 1s core levels.}
\label{fig:XPS}
\end{centering}
\end{figure}

Core level XPS measurements of the \cobaltite\ films after growth
and after annealing were carried out to assess the film
stoichiometry. The XPS spectra were obtained using the Mg K$_\alpha$
line ($h\nu = 1253.6$ eV) of a double anode x-ray source and a
double pass cylinder mirror analyzer (PHI 15-255G) set at a pass
energy of 25 eV (energy resolution of about 0.8 eV). XPS spectra of
the O 1s and Co 2p lines of the as-grown and annealed films are
shown in Fig.~\ref{fig:XPS}. Corrections to the data include a
five-point adjacent smoothing, x-ray satellite correction and
correction of energy shifts due to charging (aligned with respect to
the Co 2p peaks, using the energy assignments given in
\cite{CBR76,HU77}). One observation is that the Co 2p spectra for
both samples are very similar, showing that no significant changes
in stoichiometry or in the ionic state of the Co cations occur as a
consequence of annealing. A second observation is that the Co 2p
spectra are characteristic of a \cobaltite\ ionic environment
\cite{CBR76,HU77}, with strongly suppressed shake-up peaks compared
to those of CoO
\cite{BGD75,CBR76,HU77,OS92a,LAC+99,WHMS04,WAH08,PMCL08}. The O 1s
photoemission line is also similar before and after annealing. The
additional shoulder observed at higher binding energies has been
attributed to adsorbed oxygen
\cite{CBR76,JD79,KBR+84,KGBW93,GKBW95,JFEG95,CNL96,PMCL08} or to
surface hydroxylation \cite{HU77,PMCL08}; for the annealed sample
the shoulder is much more pronounced compared to the as-grown film,
consistent with surface hydroxylation through water adsorption upon
exposure to air.

\begin{figure}[t!bh]
\begin{centering}
\includegraphics*[width=\columnwidth]{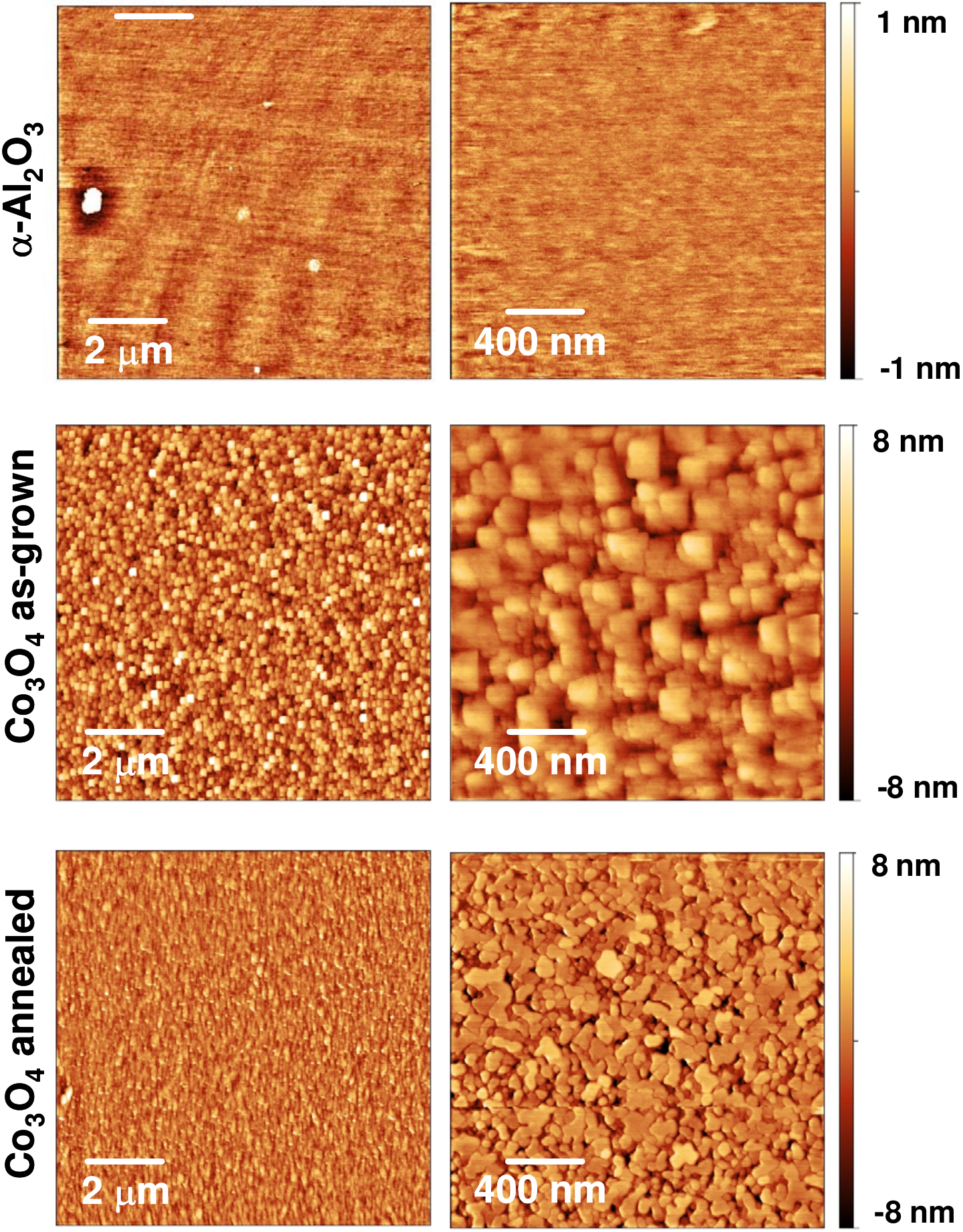}
\caption{Atomic force microscopy (AFM) scans of (a) the
\sapphire(0001) substrate surface as received, (b) the as-grown and
(c) annealed film surface for the 38 nm \cobaltite\ sample.}
\label{fig:AFM}
\end{centering}
\end{figure}

The surface morphology was also probed {\it ex situ} using atomic
force microscopy (AFM) in contact mode at room temperature. The AFM
data were corrected and analyzed using the Gwyddion freeware package
\cite{Gwyddion}; corrections to the data included planarisation,
background-correction and removal of faulty scan lines. Typical AFM
surface morphology profiles for the 38 nm sample are shown in
Fig.~\ref{fig:AFM} for $10\times 10$ $\mu$m$^2$ and $2\times 2$
$\mu$m$^2$ scanning areas. The surface of the \sapphire\ substrate
is atomically flat (average roughness of 0.2 nm), in agreement with
the RHEED and LEED results. The surface profile of the as-grown
\cobaltite\ film reveals a pronounced surface texture (rms roughness
parameter of 2.4 nm), with a characteristic in-plane correlation
length of about 80 nm determined from the $10 \times 10$ $\mu$m
scans. Although regular in-plane and square in shape, these islands
are not flat, in agreement with the spotty LEED and
transmission-like RHEED patterns. Annealing leads to a significant
change in the surface morphology, including a meandering island
shape and a significant flattening of the islands, resulting in very
uniform contrast in the higher magnification images. Overall, the
surface is more uniform for the annealed film (rms roughness
amplitude of 1.5 nm), although the island size is reduced in the
process (giving an in-plane correlation length of $60\pm 20$ nm for
the annealed surface). The observation in AFM of smoother surfaces
for the annealed film agree with the streakier RHEED patterns
observed for the annealed films, and contrast with the more 3D
transmission-like patterns of the as-grown films.

\begin{figure}[t!bh]
\begin{centering}
\includegraphics*[width=\columnwidth]{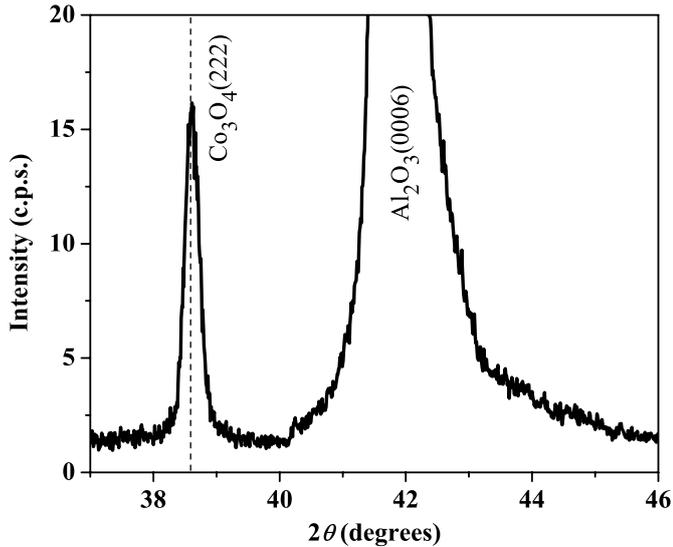}
\caption{X-ray diffraction ($\theta-2\theta$ scan) of the annealed
sample (38 nm \cobaltite) near the (0006) diffraction plane of
\sapphire, showing the \cobaltite(222) diffraction peak. Dashed line
indicates the position of the bulk (222) diffraction peak of
\cobaltite.} \label{fig:XRD}
\end{centering}
\end{figure}

Structural characterization of the \cobaltite\ films was carried out
{\it ex situ} by x-ray scattering measurements on a Shidmazu
diffractometer using the Cu K$_\alpha$ line
($\lambda_{\mathrm{K}\alpha 1} = 1.540606$ \AA) with a Ni filter to
remove the Cu K$_\beta$ lines. Comparison with \sapphire\ XRD
spectra allows one to determine readily the diffraction peaks
originating from the \cobaltite\ films (not shown). The XRD results
show that the spectra for the as-grown and annealed films are
identical in the range from $2\theta = 20\mbox{ - }100^\mathrm{o}$,
with diffraction peaks that coincide with the (hhh) planes ($h=
1,2,3,4$) of \cobaltite, at $2\theta$ values corresponding to the
bulk lattice parameter. This confirms the spinel crystalline phase
and epitaxy of the \cobaltite\ film, and also that the films are
fully relaxed both after growth and after annealing. In
Fig.~\ref{fig:XRD} we show a detail of the XRD spectra of the
annealed film around the (0006) \sapphire\ peak, showing the (222)
\cobaltite\ diffraction peak.

\begin{figure}[t!bh]
\begin{centering}
\includegraphics*[width=\columnwidth]{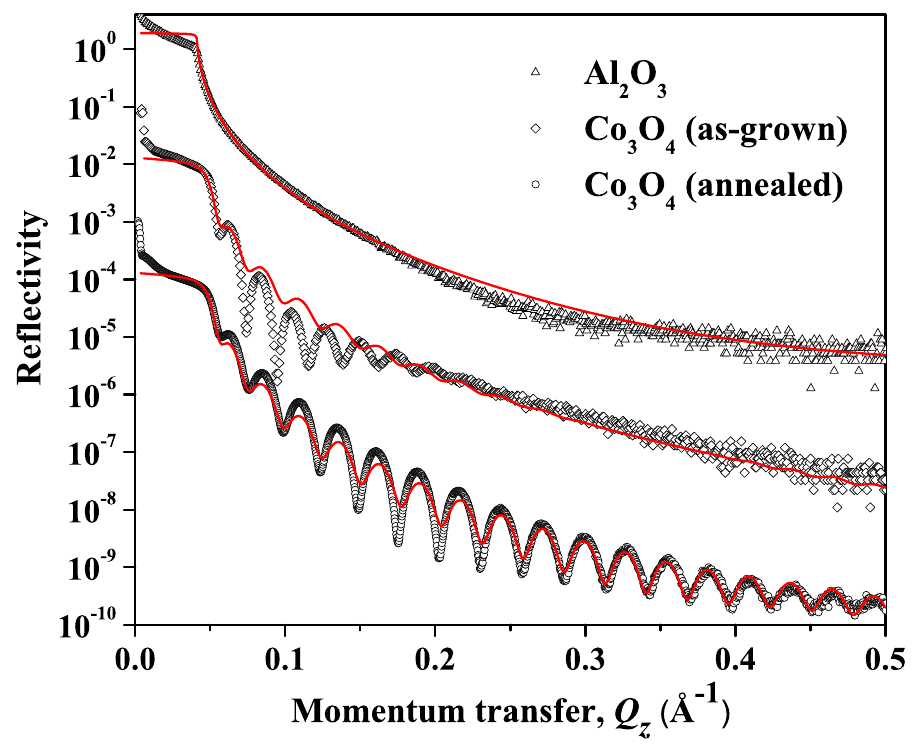}
\caption{X-ray reflectivity spectra (symbols) for the \sapphire\
substrate and for the as-grown and annealed 22 nm \cobaltite\ film.
Data have been shifted vertically by a factor of 100 for convenient
data display; solid lines are fits to the data. (Data have been
corrected for the `fingerprint' effect at low angles of incidence.)}
\label{fig:XRR}
\end{centering}
\end{figure}

The XRR spectra of the substrate, as-grown, and annealed 22 nm
\cobaltite\ films are shown in Fig.~\ref{fig:XRR}. The spectrum of
the substrate (as-received) allows one to determine its surface
roughness, which is estimated as 1 nm from the fit to the data (fits
are shown as solid lines in Fig.~\ref{fig:XRR}). The reflectivity
spectrum for the as-grown films show a rapid drop in the Kiessig
fringes' amplitude with momentum transfer; considering a single
\cobaltite\ layer with bulk scattering density does not yield a good
fit to the data, suggesting the presence of a non-uniform or graded
interface layer. In contrast, the reflectivity spectrum of the
annealed films show oscillations over the entire momentum transfer
range probed, indicating a significant improvement of the interface
sharpness. Fits to the data indicate that the changes occur most
significantly at the \sapphire\ interface, where the roughness
amplitude is found to decrease from 2.0 to 0.6 nm for the 22 nm
sample, and from 4.0 to 1.5 nm for the 38 nm \cobaltite\ film.
Hence, we interpret the strongly damped XRR spectra of the as-grown
films as resulting from chemical or structural disorder at the
\cobaltite/\sapphire\ interface, as also suggested from the RHEED
pattern evolution during growth, which shows the \sapphire\ sharp
diffraction spots becoming streaky at about 15 \AA\ and broadening
at about 30 \AA. In XRR, interface roughness may originate from
graded interfaces, interdiffusion, and morphological roughness
proper \cite{WPH+96}. The improvement in interface sharpness upon
annealing also suggests that the interface roughness in the as-grown
film does not result from a reacted interface region, which should
increase in thickness with annealing. Fits to the data give film
thicknesses of 22.5 nm and 38.5 nm for the two samples, in good
agreement with the nominal thicknesses estimated from the Co
evaporation rate and the ratio between the \cobaltite\ and Co mass
densities. The rms surface roughness of the \cobaltite\ films is
found to increase slightly upon annealing from 1 to 2 nm for the 22
nm film, and from 2 to 2.5 nm for the 38 nm sample. The accuracy of
the fits are within $\pm 0.5$ nm for the interface roughness and
$\pm 0.2$ nm for the thickness.

These results are in marked contrast with those obtained for
\cobaltite(110)/\spinel(110) films, where annealing under similar
conditions results in atomically flat surfaces, associated with one
particular termination of \cobaltite\ along the [110] direction
\cite{VWA+09}. Two factors can be expected to give rise to this
difference in behaviour: the relative surface stability of the (110)
and (111) crystal planes, and the role of strain in determining the
surface morphology and surface roughness in particular. Surface
roughening starts at the very early stages of the \cobaltite(111)
growth, and while this could be driven by misfit strain relaxation,
we find that a similar roughening process also occurs in
\cobaltite/\spinel(110), where the lattice mismatch is virtually
zero \cite{VWA+09}. Transmission electron microscopy results for the
\cobaltite/\spinel(110) interface show sharp and relatively well
defined interfaces \cite{VWA+09}, suggesting that the disorder
observed in the XRR data could be due to modified film stoichiometry
at the interface, leading to a graded interface structure; for the
\cobaltite/\spinel(110) system the starting growth temperature was
set to 770 K for the first 1 nm interface layer to optimize
crystalline growth, which could lead to variations in interface
composition and to a degree of interface grading. In the present
study, the growth temperature was constant throughout the film
growth; however, differences in the substrate and film crystal
structure may favor an initial cationic arrangement closer to the
corundum structure than to the spinel; for example, if  the Co in
the first few layers occupy an excess of octahedral sites, the film
stoichiometry could be closer to Co$_2$O$_3$, resulting in a graded
interface. This also suggests that film roughening may arise from a
kinetic process common to both \cobaltite/\sapphire(0001) and
\cobaltite/\spinel(110) systems, for example, induced by limitations
in atomic surface diffusion during growth. At annealing temperatures
of about 800 K, atomic diffusion is sufficient to promote surface
ordering in the \cobaltite\ films, but while the (110) films become
atomically smooth, the (111) films retain their pronounced island
structure and are not atomically flat, as indicated from the AFM,
XRR and RHEED results. This points to a lower stability of the (111)
surface, at least under the annealing conditions.

Another intriguing aspect of this work is the fact that the polar
\cobaltite(111) surfaces, for both the as-grown and annealed films,
are ($1\times 1$) \cite{MBG+08}, a finding similar to that found for
Fe$_3$O$_3$(111) (depending on the growth and surface preparation
conditions) \cite{GKCB97,RW99,SBMM03} and \cobaltite(110) film
surfaces \cite{VWA+09}. For Fe$_3$O$_3$(111) films grown on Pt(111),
the ($1\times 1$) pattern was attributed to a surface structure
consisting of an Fe-terminated layer, with large relaxations of the
top four atomic layers, as determined from quantitative LEED
intensity analysis \cite{RW99}; such atomic relaxations are expected
to induce strong modifications in the surface electron density of
states, which could lead to charge compensation and to surface
stability \cite{RW99,Noguera00}. The results for the \cobaltite\
surfaces have been explained in terms of a charge compensation
mechanism consisting of a surface spinel inversion process, whereby
tetrahedral Co$^{2+}$ cations revert to a trivalent state
\cite{MBG+08,VWA+09}, leading to stable ($1 \times 1$) surfaces. We
envisage that a similar mechanism may be in play for the (111)
surface.

\section{Conclusions}

In summary, we have demonstrated epitaxial growth of \cobaltite(111)
thin films on \sapphire(0001) by oxygen assisted molecular beam
epitaxy. The films grow with the \cobaltite\ stoichiometry, which
remains unchanged upon post-growth annealing. However, the interface
and surface structure of the \cobaltite(111) film changes
considerably upon annealing. Annealing leads to a significant
improvement of the \cobaltite/\sapphire\ interface structure, while
the \cobaltite/air interface is also strongly modified. Unlike
\cobaltite/\spinel(110), annealing does not lead to atomically
smooth surfaces in \cobaltite/\sapphire(0001). Stable ($1 \times 1$)
surfaces are observed in both as-grown and annealed films, which is
explained by a surface spinel inversion process that leads to
surface charge compensation.

\section*{Acknowledgements}

The authors acknowledge financial support by the NSF through MRSEC
DMR 0520495 (CRISP), MRSEC DMR 0705799, the ONR (C.H.A.), and the
U.S. Department of Energy Basic Energy Sciences Grant Numbers
DEFG02-98ER14882 and DE-FG02-06ER15834 (E.I.A.).


\begin{thebibliography}{10}
\expandafter\ifx\csname url\endcsname\relax
  \def\url#1{\texttt{#1}}\fi
\expandafter\ifx\csname urlprefix\endcsname\relax\def\urlprefix{URL
}\fi

\bibitem{PBBC80}
J.~P. Picard, G.~Baud, J.~P. Besse, R.~Chevalier, Journal of the
Less-Common
  Metals 75 (1980) 99.

\bibitem{CJM71}
J.~Chenavas, J.~C. Joubert, M.~Marezio, Solid State Commun. 9 (1971)
1057.

\bibitem{Samsonov82}
G.~V. Samsonov (Ed.), The oxide handbook, 2nd Edition, IFI/Plenum,
New York,
  1982.

\bibitem{CS97}
M.~Catti, G.~Sandrone, Faraday Discuss. 106 (1997) 189.

\bibitem{Tasker79}
P.~W. Tasker, J. Phys. C: Solid State Phys. 12 (1979) 4977.

\bibitem{Noguera00}
C.~Noguera, J. Phys.: Condens. Matter 12 (2000) R367.

\bibitem{GFN08}
J.~Goniakowski, F.~Finocchi, C.~Noguera, Rep. Prog. Phys. 71 (2008)
016501.

\bibitem{HKWG78}
W.~A. Harrison, E.~A. Kraut, J.~R. Waldrop, R.~W. Grant, Phys. Rev.
B 18 (1978)
  4402.

\bibitem{JPS+02}
M.~Gajdardziska-Josifovska, R.~Plass, M.~A. Schofield, D.~R. Giese,
R.~Sharma,
  J. Electron Microsc. 51 (2002) S13.

\bibitem{LPP+05}
V.~K. Lazarov, R.~Plass, H.-C. Poon, D.~K. Saldin, M.~Weinert, S.~A.
Chambers,
  M.~Gajdardziska-Josifovska, Phys. Rev. B 71 (2005) 115434.

\bibitem{VWA+09}
C.~A.~F. Vaz, H.~Wang, C.~H. Ahn, V.~E. Henrich, M.~Z. Baykara,
  T.~Schwendemann, N.~Pilet, B.~J. Albers, U.~Schwarz, L.~H. Zhang, Y.~Zhu,
  J.~Wang, E.~I. Altman, Surf. Sci. 603 (2009) 291.

\bibitem{MBG+08}
W.~Meyer, K.~Biedermann, M.~Gubo, L.~Hammer, K.~Heinz, J. Phys.:
Condens.
  Matter 20 (2008) 265011.

\bibitem{PMCL08}
S.~C. Petitto, E.~M. Marsh, G.~A. Carson, M.~A. Langell, J.
Molecular Catalysis
  A: Chemical 281 (2008) 49.

\bibitem{TLH08}
X.~Tang, J.~Li, J.~Hao, Materials Research Bulletin 43 (2008) 2912.

\bibitem{FPW00}
C.~M. Fang, S.~C. Parker, G.~de~With, J. Am. Ceram. Soc. 83 (2000)
2082.

\bibitem{DPW94}
M.~J. Davies, S.~P. Parker, G.~W. Watson, J. Mater. Chem. 4 (1994)
813.

\bibitem{SB80}
R.~L. Stewart, R.~C. Bradt, J. Mater. Sci. 15 (1980) 67.

\bibitem{Bradt97}
R.~C. Bradt, Cleavage of ceramic and mineral single crystals, in:
K.~S. Chan
  (Ed.), George R. Irvin Symposium on Cleavage Fracture, Warrendale, PA, 1997,
  p. 355.

\bibitem{FWP01}
C.~M. Fang, G.~de~With, S.~C. Parker, J. Am. Ceram. Soc. 84 (2001)
1553.

\bibitem{LDL+05}
N.~J. van~der Laag, A.~J.~M. van Dijk, N.~Lousberg, G.~de~With,
L.~J. M.~G.
  Dortmans, J. Am. Ceram. Soc. 88 (2005) 660.

\bibitem{NH62}
R.~E. Newnham, Y.~M.~D. Haan, Zeitschrift f{\"u}r Kristallographie
117 (1962)
  235.

\bibitem{Cousins81}
C.~S.~G. Cousins, J. Phys. C: Solid State Phys. 14 (1981) 1585.

\bibitem{KE90}
A.~Kirfel, K.~Eichhorn, Acta Cryst. A 46 (1990) 271.

\bibitem{Wyckoff64}
R.~W.~G. Wyckoff, Crystal structures, 2nd Edition, Vol.~2,
Interscience
  Publishers, New York, 1964.

\bibitem{GEL92}
J.~Guo, D.~E. Ellis, D.~J. Lam, Phys. Rev. B 45 (1992) 13647.

\bibitem{GL94}
T.~J. Godin, P.~LaFemina, Phys. Rev. B 49 (1994) 7691.

\bibitem{MVG93}
I.~Manassidis, A.~D. Vita, M.~Gillan, Surf. Sci. Lett. 285 (1993)
L517.

\bibitem{AR97}
J.~Ahn, J.~Rabalais, Surf. Sci. 388 (1997) 121.

\bibitem{GRBG97}
P.~Gu{\'e}nard, G.~Renaud, A.~Barbier, M.~Gautier-Soyer, Surf. Rev.
Lett. 5
  (1997) 321.

\bibitem{Renaud98}
G.~Renaud, Surface Science Reports 32 (1998) 1.

\bibitem{JC01}
E.~A.~A. Jarvis, E.~A. Carter, J. Phys. Chem. B 105 (2001) 4045.

\bibitem{WAM+06}
E.~Wallin, J.~M. Andersson, E.~P. M{\"u}nger, V.~Chirita,
U.~Helmersson, Phys.
  Rev. B 74 (2006) 125409.

\bibitem{Chang68}
C.~C. Chang, J. Appl. Phys. 39 (1990) 5570.

\bibitem{FS70}
T.~M. French, G.~A. Somorjai, J. Appl. Phys. 74 (1970) 2489.

\bibitem{HC94}
V.~E. Henrich, P.~A. Cox, The surface science of metal oxides,
Cambridge
  University Press, Cambridge, 1994.

\bibitem{SHOS99}
T.~Suzuki, S.~Hishita, K.~Oyoshi, R.~Souda, Surf. Sci. 437 (1999)
289.

\bibitem{WMSH00}
C.~F. Walters, K.~F. McCarty, E.~A. Soares, M.~A.~Van Hove, Surf.
Sci. 464
  (2000) L732.

\bibitem{BR01}
C.~Barth, M.~Reichling, Nature 414 (2001) 54.

\bibitem{SHWM02}
E.~A. Soares, M.~A.~Van Hove, C.~F. Walters, K.~F. McCarty, Phys.
Rev. B 65
  (2002) 195405.

\bibitem{MP04}
A.~Marmier, S.~C. Parker, Phys. Rev. B 69 (2004) 115409.

\bibitem{Kelber07}
J.~A. Kelber, Surf. Sci. Rep. 62 (2007) 271.

\bibitem{GKA05}
W.~Gao, R.~Klie, E.~Altman, Thin Solid Films 485 (2005) 115.

\bibitem{AWN+00}
R.~Anton, T.~Wiegner, W.~Naumann, M.~Liebmann, C.~Klein, C.~Bradley,
Rev. Sci.
  Instrum. 71 (2000) 1177.

\bibitem{TK55}
H.~P. Tripp, B.~W. King, J. American Ceramic Soc. 38 (1955) 432.

\bibitem{KY81a}
K.~Koumoto, H.~Yanagida, Jpn. J. Appl. Phys. 20 (1981) 445.

\bibitem{OS92a}
M.~Oku, Y.~Sato, Appl. Surf. Sci. 55 (1992) 37.

\bibitem{RF63}
R.~C. Rossi, R.~M. Fulrath, J. Am. Ceram. Soc. 46 (1963) 145.

\bibitem{LCC05}
C.-M. Liu, J.-C. Chen, C.-J. Chen, J. Crystal Growth 285 (2005) 275.

\bibitem{GKCB97}
Y.~Gao, Y.~J. Kim, S.~A. Chambers, G.~Bai, J. Vac. Sci. Technol. A
15 (1997)
  332.

\bibitem{Stowell75}
M.~J. Stowell, Defects in epitaxial deposits, in: J.~W. Matthews
(Ed.),
  Epitaxial growth, Part B, Academic Press, Inc., 1975, p. 437.

\bibitem{CBR76}
T.~J. Chuang, C.~R. Brundle, D.~W. Rice, Surf. Sci. 59 (1976) 413.

\bibitem{HU77}
J.~Haber, L.~Ungier, J. Electron Spectrosc. Relat. Phenom. 12 (1977)
305.

\bibitem{BGD75}
J.~P. Bonnelle, J.~Grimblot, A.~D'Huysser, J. Electron Spectrosc.
Relat.
  Phenom. 7 (1975) 151.

\bibitem{LAC+99}
M.~A. Langell, M.~D. Anderson, G.~A. Carson, L.~Peng, S.~Smith,
Phys. Rev. B 59
  (1999) 4791.

\bibitem{WHMS04}
H.~A. Hagelin-Weaver, G.~B. Hoflund, D.~M. Minahan, G.~N. Salaita,
Appl. Surf.
  Sci. 235 (2004) 420.

\bibitem{WAH08}
H.-Q. Wang, E.~I. Altman, V.~E. Henrich, Phys. Rev. B 77 (2008)
085313.

\bibitem{JD79}
Y.~Jugnet, T.~M. Duc, J. Phys. Chem. Solids 40 (1979) 29.

\bibitem{KBR+84}
Y.~M. Kolotyrkin, I.~D. Belova, Y.~E. Roginskaya, V.~B. Kozhevnikov,
D.~S.
  Zakhar'in, Y.~N. Venevtsev, Materials Chemistry and Physics 11 (1984) 29.

\bibitem{KGBW93}
B.~Klingenberg, F.~Grellner, D.~Borgmann, G.~Wedler, Surf. Sci. 296
(1993) 374.

\bibitem{GKBW95}
B.~Klingenberg, F.~Grellner, D.~Borgmann, G.~Wedler, J. Electron
Spectrosc.
  Relat. Phenom. 71 (1995) 107.

\bibitem{JFEG95}
V.~M. Jim{\'e}nez, A.~Fern{\'a}ndez, J.~P. Espin{\'o}s, A.~R.
  Gonz{\'a}lez-Elipe, J. Electron Spectrosc. Relat. Phenom. 71 (1995) 61.

\bibitem{CNL96}
G.~A. Carson, M.~H. Nassir, M.~A. Langell, J. Vac. Sci. Technol. A
14 (1996)
  1637.

\bibitem{Gwyddion}
{h}ttp://gwyddion.net

\bibitem{WPH+96}
M.~Wormington, I.~Pape, T.~P.~A. Hase, B.~K. Tanner, D.~K. Bowen,
Phil. Mag.
  Lett. 74 (1996) 211.

\bibitem{RW99}
M.~Ritter, W.~Weiss, Surf. Sci. 432 (1999) 81.

\bibitem{SBMM03}
I.~V. Shvets, N.~Berdunov, G.~Mariotto, S.~Murphy, Europhys. Lett.
63 (2003)
  867.

\end{thebibliography}



\end{document}